\documentclass[12pt]{article}
\usepackage{fleqn}

\newenvironment{my_hacked_eqn}[1]
	{\begin{eqnarray}\label{#1}}
	{\end{eqnarray}}
\newcommand{\mybibitem}[1]{\bibitem{#1}}

\newcommand{\be}[1]{\begin{my_hacked_eqn}{#1}}
\newcommand{\eq}{\end{my_hacked_eqn}}

\def\vare{\varepsilon}

\def\a{\alpha}
\def\b{\beta}
\def\c{\chi}
\def\d{\delta}
\def\f{\phi}	

\def\m{\mu}
\def\n{\nu}

\def\p{\pi}			
\def\s{\sigma}			

\def\F{\Phi}

\def\ca{\mathcal{A}}
\def\cb{\mathcal{B}}

\def\cd{\mathcal{D}}

\def\cf{\mathcal{F}}
\def\cg{\mathcal{G}}

\def\ck{\mathcal{K}}
\def\cl{\mathcal{L}}
\def\cm{\mathcal{M}}

\def\cs{\mathcal{S}}

\def\cz{\mathcal{Z}}

\def\pa{\partial}			
\def\bpa{\bar{\partial}}	 
\def\de{\nabla}				
\def\half{\frac{1}{2}}

\newcommand{\NPB}[1]{Nucl.\ Phys.\ \textbf{B#1}}
\newcommand{\PLB}[1]{Phys.\ Lett.\ \textbf{B#1}}
\newcommand{\CMP}[1]{Comm.\ Math.\ Phys.\ \textbf{#1}}

\newcommand{\xxx}[1]{\mbox{hep-th/\textbf{#1}}}

\def\MR{M.~Ro\v{c}ek {} }
\def\AGMR{A.~Giveon, \MR}

\makeatletter
\@addtoreset{equation}{section}
\@addtoreset{equation}{subsection}
\def\theequation{\ifnum\value{section}=0 \arabic{equation}\ignorespaces
\else \ifnum\value{subsection}=0 \thesection.\arabic{equation}\ignorespaces
\else \thesection.\arabic{subsection}.\arabic{equation}\ignorespaces
                       \fi
                 \fi}
\@addtoreset{table}{section}
\@addtoreset{table}{subsection}
\def\thetable{\ifnum\value{section}=0 \arabic{table}\ignorespaces
\else \ifnum\value{subsection}=0 \thesection.\arabic{table}\ignorespaces
\else \thesection.\arabic{subsection}.\arabic{table}\ignorespaces
                    \fi
              \fi}
\makeatother

\def\bK{\bar{K}}

\def\Lra{\Longrightarrow}

\begin{document}

\begin{titlepage}

\begin{flushleft}
ITP-SB-95-50 \\
USITP-95-11 \\
hep-th/9512025 \\
May 23, 1996
\end{flushleft}

\begin{center}

\vskip3em
\textbf{\huge Poisson-Lie T-Duality: the Path-Integral Derivation}

\vskip3em
{\Large \textsl{Eugene Tyurin\hskip1em{\rm and}\hskip1em
Rikard von Unge\footnote{On
leave from Department of Physics, Stockholm University, Sweden}}}
\vskip1em

\textsl{ Institute for Theoretical Physics, SUNY at Stony Brook, \\
Stony Brook, NY 11794-3840, USA \\ \vskip1ex E-mail: \hskip1ex
gene {\rm and} unge@insti.physics.sunysb.edu }

\end{center}

\vfill

\begin{abstract}

\noindent
We formulate Poisson-Lie T-duality in a path-integral manner that
allows us to analyze the quantum corrections.  Using the
path-integral, we rederive the most general form of a Poisson-Lie
dualizeable background and the generalized Buscher transformation
rules it has to satisfy.

\end{abstract}

\vfill\hfill {\footnotesize Typeset by \LaTeXe}

\end{titlepage}

\section{Introduction}\label{intro}

The motive that fuels diverse research effort in the field of duality is
the hope that the ``great unification'' of String Theory and low-energy
physics exists.  Dualities (of different kinds) are a product of String
Theory.  As such, ultimately they should be able to allow us to link the
properties of space-time with the properties of the particle spectrum
through the appropriate choice of compactification.

Speaking less ambitiously, target space dualities provide us with
a prescription how to identify apparently different target space
backgrounds as equivalent or, to put it differently, how to classify
physically inequivalent string vacua and, thus, access the moduli space
of String Theory.

Dualities relate different regimes of the theory.  Target space duality
was originally discovered in toroidal compactifications where it changed
our understanding of the concept of distance.  Namely, as it turns out,
there is a minimal (self-dual) distance $R_0=R_0^{-1}$ in String Theory.
Any smaller distance $R<R_0$ is equivalent to $R^\prime = R^{-1} > R_0$.

Recently, Klim\v{c}ik and \v{S}evera proposed a generalization of
abelian and traditional non-abelian dualities called Poisson-Lie
T-duality \cite{sever}.  They classified all Poisson-Lie dualizeable
backgrounds and proved the phase space equivalence of a model and
its Poisson-Lie dual. Later, a path-integral argument was given
\cite{AKS} for the very special case when the $\s$-model admits chiral
equations of motion.  We present a completely general, unambiguous
path-integral scheme that allows one to handle Poisson-Lie T-duality
and that reduces to the established ways of treating ``older'' target
space dualities.

As duality research progressed towards complicated setups yielding
sometimes quite unexpected results, it became more and more important to
clearly realize what does one mean by ``equivalence'' in duality.

In the simplest case, backgrounds with abelian isometries (toroidally
compactified theories) were proven to be equivalent Conformal
Field Theories, at least in the lowest order of the string coupling
$\a^\prime$.  Traditional non-abelian duals were proven to be
conformal (except for non-semisimple duality groups) \cite{AAGY,conf},
with the relation between those CFTs as that of an orbifold by a
continuous group \cite{AGMR}.

As we intend to demonstrate in Section \ref{path}, non-abelian dual
backgrounds (in both traditional and Poisson-Lie T-dualities) are
equivalent in the same sense as in the case of abelian duality modulo
global issues\footnote{We want to stress that global issues are not
addressed in this paper. This paper achieves for Poisson-Lie T-duality
what the second paper of \cite{bush} achieved for abelian T-duality.}
\cite{dqc} (except for a contribution to the trace anomaly if the
structure constants are not traceless \cite{AGMR,AAGY,conf}). Our
path-integral procedure also shows that the dual models have to
incorporate certain quantum corrections of their respective classical
target space backgrounds (dilaton shifts in both models as opposed to
only one of them).

The outline of the paper is as follows. We start by explaining
the original (algebraic) framework of Poisson-Lie target space
duality due to Klim\v{c}ik and \v{S}evera \cite{sever} that
states that dualizeable backgrounds must satisfy a certain
system of partial differential equations (\ref{lieeqn}),
(\ref{duallieeqn}) and an extra condition (\ref{match}). Next, in
Section \ref{gensolution} we construct the most general background
that solves (\ref{lieeqn},~\ref{duallieeqn},~\ref{match}).  Using
the result of Section~\ref{gensolution}, we introduce the new
path-integral formulation (\ref{WZWonD}) of Poisson-Lie T-duality in
Section~\ref{path}.  We perform the analysis of quantum corrections in
the path-integral in Section~\ref{quantum} and incorporate spectator
fields in the Section~\ref{spectators} for completeness. We use the
actions we obtained with our method to derive the generalized Buscher
rules (\ref{Brules}) for Poisson-Lie duality.  We also make thorough
comparison of all of our results with the abelian and traditional
non-abelian dualities.

\section{Basics of Poisson-Lie T-Duality}\label{basics}

In this section we introduce the scheme of Poisson-Lie T-duality which was
recently proposed by Klim\v{c}ik and \v{S}evera \cite{sever}.  Within
this new formalism, $\s$-models dual to each other are treated on equal
footing and the existence of inverse duality transformations (which was
lacking in ``traditional'' non-abelian duality \cite{que,AGMR}) is no
longer a problem.

The central idea of \cite{sever} was to abandon the requirement
that a $\s$-model has a set of isometries (and corresponding to
them conservation laws) in favor of the so-called ``non-commutative
conservation laws'' which can be thought of as non-abelian
generalization of the usual conservation laws.

Suppose we are given a $\s$-model defined on some manifold $\cm$:
\be{model}
\cs[x] = \int d^2 z\: \pa x^i F_{ij}(x) \bpa x^j
\eq
We can define the left (right) group action as
\be{varx}
\d x^i = \vare^a v^i_a,
\eq
where ${v^i}_a$ are correspondingly right (left) invariant
frames in the Lie algebra $\cg$ of the group G.  The reasoning is as
follows:
\be{leftright}
\d g=\vare g \Lra \vare =\d g g^{-1} = e^R_i\d x^i\quad\mbox{ or }\quad
\d x^i = \vare^{a}v^{i}_{R\:a} ,\quad v=e^{-1}.
\eq
This variation
generates the currents\footnote{When $F_{ij}$ has a Wess-Zumino term,
there may be corrections to (\ref{currents}) as discussed in \cite{myko}.}
\be{currents}
K_a=\pa x^i F_{ij} v^j_a \qquad \bK_a=v^i_a F_{ij}\bpa x^j
\eq
and the current 1-form
\be{curform}
\ck_a=K_a dz+{\bK}_a d\bar z .
\eq
The total variation of (\ref{model}) under (\ref{varx}) equals:
\be{varaction}
\d\cs = \int d\vare^a\wedge \star\ck_a +
\int d^2 z\: \pa x^i\bpa x^j \vare^a\cl_{v_a}F_{ij}\:,
\eq
where $\cl$ denotes a Lie derivative, and $\star$ is a Hodge star.

We now say that the given model (\ref{model}) admits ``non-commutative
conservation laws'' if the following equation\footnote{Note that the
usual conservation law is simply $d\star\ck = 0$ which corresponds to
$\tilde{f}=0$ in (\ref{dstar}).} holds on-shell: 
\be{dstar}
d\star\ck_a + \half\tilde{f}_a^{bc}\star\ck_b \wedge \star\ck_c = 0.
\eq
The above equations of motion are automatically satisfied if
$\,\star\ck$ is a pure gauge field in certain ``dual'' algebra
$\tilde{\cg}$:
\be{dual}
\star\ck_a \tilde{T}^a\equiv\star\ck =\tilde{g}^{-1}d\tilde g\in\tilde{\cg},
\qquad \tilde g\in\tilde G, \qquad
\left[ \tilde  T^b, \tilde T^c \right] = \tilde{f}_a^{bc}\tilde T^a
\eq
This is the general feature of duality: equation of motion becomes a
Bianchi identity when expressed in terms of dual fields.

Note that we have called (\ref{dstar}) ``equations of motion'': we have
already implied that we consider only $\s$-models with the following
property of their backgrounds $F_{ij}$ (c.f. (\ref{currents})):
\be{lieeqn}
\cl_{v_a}F_{ij} = F_{ik}v^k_b \tilde{f}_a^{bc} v^l_c F_{lj}.
\eq
Equation (\ref{lieeqn}) is the central statement of the
Klim\v{c}ik-\v{S}evera approach, because it is a
geometrical, non-dynamical equation for the background field $F_{ij}$.
By demanding the closure of (\ref{lieeqn})
( $\left[ \cl_{v_a}, \cl_{v_b} \right] = f^c_{ab}\cl_{v_c}$),
we obtain the following consistency condition:
\be{bialg}
f_{dc}^a \tilde{f}_a^{rs} = \tilde{f}_c^{as} f_{da}^r +
\tilde{f}_c^{ra} f_{da}^s - \tilde{f}_d^{as} f_{ca}^r -
\tilde{f}_d^{ra} f_{ca}^s,
\eq
which is known in mathematics to be the relation for the structure
constants of the Lie bi-algebra $(\cg,\tilde{\cg})$ (for a review, see e.g.
\cite{math}).

It is now natural to postulate that the dual to (\ref{model}) $\s$-model
should obey the same condition as (\ref{lieeqn}) but with the tilded and
un-tilded variables interchanged:
\be{duallieeqn}
\cl_{\tilde v^a}\tilde{F}^{ij} = \tilde{F}^{ik}\tilde{v}_k^b f^a_{bc}
\tilde{v}_l^c \tilde{F}^{lj}.
\eq
and the backgrounds related by
\be{match}
\left( F(x=0)\right)_{ij}^{-1} = \tilde{F}^{ij}(\tilde x = 0).
\eq

It should be noted that the dilaton contribution cannot
be estimated within this approach \cite{sever}.

As we have mentioned above, a new group structure, called Drinfeld double,
emerges in Poisson-Lie T-duality. The Drinfeld double $D$ comes equipped
with an inner product (non-degenerate symmetric bi-linear form) that
is invariant under the adjoint action of the full double, and has the
following properties:
\be{innerprod}
\langle T_a, T_b \rangle = 0 \qquad
\langle \tilde{T}^a, \tilde{T}^b \rangle = 0 \qquad
\langle T_a, \tilde{T}^b \rangle = \d_a^b\: ,
\eq
\be{adinvar}
\langle l\:T^A l^{-1}, T^B\rangle = \langle T^A, l^{-1}T^B l\rangle\:
\qquad l\in D,
\eq
where $T_a\in\cg$, $\tilde{T}^a\in\tilde{\cg}$ and $T^A\in\cd$.
Also note that any $l\in D$ can be uniquely decomposed (at least
close to the identity):
\be{decompose}
l = \tilde{h}g = h\tilde{g}\qquad\mbox{ where }
g,h \in G\quad \tilde{g},\tilde{h} \in\tilde{G}
\eq
Let us also introduce some notation used later in the paper:
\be{notation}
\m^{ab}(g) = \langle g\:\tilde{T}^a\, g^{-1},\tilde{T}^b\rangle \qquad
\n^a_b(g) = \langle g\:\tilde{T}^a\, g^{-1}, T_b\rangle \nonumber\\
\a^a_b(\tilde{g})
  = \langle \tilde{g}\:T_b\,\tilde{g}^{-1}, \tilde{T}^a\rangle \qquad
\b_{ab}(\tilde{g}) = \langle \tilde{g}\:T_a\,\tilde{g}^{-1},T_b\rangle
\eq
One may check that $\n(g^{-1})=\n^{-1}(g),\:
\a(\tilde{g}^{-1})=\a^{-1}(\tilde g),\;
\m(g^{-1})=\m^{t}(g)$ and $\b(\tilde{g}^{-1})=\b^{t}(\tilde{g})$
where $t$ stands for transpose.

This completes our overview of the algebraic approach to Poisson-Lie
target space duality.  Our next step is to find a general solution of
(\ref{lieeqn}) and then, using it as a guide, find a path-integral
formulation of Poisson-Lie target space duality.

\section{General Solution to Poisson-Lie T-Duality}\label{gensolution}

For simplicity, in this section we consider duality without spectator
fields. In other words, the target space can be identified with the
group manifold of $G$ (or $\tilde{G}$). The full problem will be
treated in section \ref{spectators}.

In traditional non-abelian duality one starts from a $\s$-model
\be{trad}
\cs_{\,\mbox{\sc tnad}} =
\int d^2 z\: (g^{-1}\pa g)^{a}E_{ab} (g^{-1}\bar{\pa}g)^{b},
\eq
where $E_{ab}$ is some constant matrix. Comparing this with
(\ref{model}) and (\ref{leftright}) one can see that
\be{TradRel}
E_{ab}=v^{i}_{La}F_{ik}v^{k}_{Lb}
\eq
is the construct one should work with in order to be able to make
a direct connection with traditional non-abelian duality.
However, the equation that was solved in \cite{sever} is
\be{KSeqn}
\cl_{v_{Rc}}(F_{ij}) =
F_{ik}v^{k}_{Ra}\tilde{f}^{ab}_{c}v^{l}_{Rb}F_{lj},
\eq
and the solution given in \cite{sever} is:
\be{KSsol}
v^{i}_{Ra}F_{ik}v^{k}_{Rb} = \left( ( \n^{-1} )_{c}^{a} +
  E^{0}_{cd}\m^{da}\right)^{-1}
  E^{0}_{ce}\n^{e}_{b},
\eq
where $E^{0}_{ab}$ is some constant matrix.

We can get the background we are interested in by multiplying the
Klim\v{c}ik-\v{S}evera solution on both sides with
\be{omega}
 e^{a}_{Ri}v^{i}_{Lb} = \langle g T_{b} g^{-1},\tilde{T}^{a}\rangle =
  \n^{-1},
\eq
which yields
\be{OurSol}
  E_{ab}(g)\equiv
  v^{i}_{La}F_{ik}v^{k}_{Lb} = \left( \left(E^{0}_{ab}\right)^{-1}+
  \m^{ac}\n_{c}^{b}\right)^{-1}.
\eq
Similarly the dual background is given by
\be{DualBG}
  \tilde{E}^{ab}(\tilde{g})\equiv
  \tilde{v}_{L}^{ai}\tilde{F}_{ik}\tilde{v}_{L}^{bk} = \left(E^{0}_{ab}+
   \b_{ac}\a^{c}_{b}\right)^{-1}.
\eq
Note that the matrices $\m\n$ and $\b\a$ are always {\em
anti-symmetric}\footnote{This follows from
$\langle\ca ,\cb\rangle =
\langle\ca, T_a\rangle\langle\tilde T^a,\cb\rangle +
\langle\ca, \tilde T^a\rangle\langle T_a,\cb\rangle$.}, in particular this
means that this duality relates backgrounds without torsion to backgrounds
with torsion.

We see that if the dual group is abelian with coordinates $\c_{a}$,
(\ref{OurSol}) and (\ref{DualBG}) reduces to
\be{SNAD}
  E_{ab}(g)&=&E^{0}_{ab},\\
  \tilde{E}^{ab}(\tilde{g})&=&\left(E^{0}_{ab}+
  \c_{c}f^{c}_{ab}\right)^{-1},
\eq
which is compatible with traditional non-abelian duality
\cite{que,AGMR}.

\section{Path-Integral Construction of Poisson-Lie T-Duality}\label{path}

Using the inner product $\langle ,\rangle$ one can write a WZW model
defined on a Drinfeld double. Since the invariant form is identically
zero on each of the sub-algebras $\cg$ and $\tilde{\cg}$, a WZW model
defined through $\langle,\rangle$ will vanish on these and therefore the
Polyakov-Wiegmann identity will have a very simple form if applied to
$I[g\tilde h]$ where $g\in G$ and $\tilde{h}\in \tilde{G}$
\be{PWident}
I[g\tilde{h}] &=& I[g]+I[\tilde{h}]+
\int d^2 z\: \langle g^{-1}\pa g,\bar{\pa}
 \tilde{h}\tilde{h}^{-1}\rangle \\ \nonumber
&=& \int d^2 z\: \langle g^{-1}\pa g,\bar{\pa}
                       \tilde{h}\tilde{h}^{-1}\rangle.
\eq
We are going to show how Poisson-Lie T-duality can be derived from
a constrained WZW-model on the double ( $l\in D$ ):
\be{WZWonD}
\cz = \int \cd l \:\d \left[\langle l^{-1}\pa l,
\tilde{T}^{a}\rangle E^{0}_{ab} -
        \langle l^{-1}\pa l,T_{b}\rangle \right] \exp (-I[l]).
\eq
While the above choice may seem strange and arbitrary, let us point out
that it is classically equivalent to another well-known model: a WZW model
with current-current interaction:
\be{classequiv}
\cs\left[l_1, l_2\right] =  I\left[ l_1 l_2 \right] -
\int d^2 z\: \langle l_1^{-1}\pa l_1, T^A \rangle E^0_{AB}
\langle T^B, \bpa l_2 l_2^{-1} \rangle\: ,
\eq
where $E^0_{AB}$ is a block-diagonal constant matrix such, that
$\left( E^0_{ab}\right)^{-1} = E^{0\:\tilde a \tilde b}$ and
${{E^0}_a}^{\tilde b} = {E^{0\:\tilde{a}}}_b = 0 $.

To obtain (\ref{WZWonD}) one should first truncate $l_2$ to lie in
$G$ (or $\tilde G$).  Then, after using the Polyakov-Wiegmann identity
(\ref{PWident}) to decompose $I[l_1 l_2]$, one finds that
$\bpa l_2 l_2^{-1}$ becomes a Lagrange multiplier enforcing the
delta-function constraint of (\ref{WZWonD}).

Unfortunately, in transition from (\ref{classequiv}) to (\ref{WZWonD})
one acquires extra Jacobians that complicate further analysis so we will
only use this observation to find the starting point for the dualization
process.

If we decompose the group element $l$ in (\ref{WZWonD}) as
$l=\tilde{h}g$, the path integral can be re-written as
\be{Gpi}
\int \cd(\tilde{h}g)\d \left[ g^{-1}\pa
  g E^{0} + \tilde{h}^{-1}\pa\tilde{h}\left(
  \m^{t} E^{0} - \n^{-1} \right)
  \right] \times \\ \nonumber
 \exp \left(-\int d^2 z\:\langle \tilde{h}^{-1}\pa\tilde{h},
\bar{\pa}g g^{-1}\rangle\right).
\eq
Since the left-invariant Haar measure splits into
\be{Haar}
 \cd(\tilde{h}g) = \cd\tilde{h}\cd g \det\left(\n^{-1}\right),
\eq
we obtain, introducing the notation
$J=g^{-1}\pa g$, $\tilde{A} = \tilde{h}^{-1}\pa \tilde{h}$
\be{Gresult}
 \int \cd g \cd\tilde{h}
    \det\left(\left(E^{0}\right)^{-1}E(g)\right)
    \d\left[\tilde{A}-
        JE\n\right]
    \exp \left(-\int d^2 z\:\tilde{A}\n^{-1} \bar{J}\right).
\eq
If we instead decompose the group element $l$ as $l = h\tilde{g}$ and
perform the analogous calculation, we get
\be{Gtilderesult}
  \int \cd h \cd\tilde{g}
    \det\left(\tilde{E}(\tilde{g})\right)
    \d\left[A-\tilde{J}\tilde{E}\a\right]
    \exp \left(-\int d^2 z\: A\a^{-1} \tilde{\bar{J}}\right),
\eq
where now $\tilde{J}=\tilde{g}^{-1}\pa \tilde{g}$, $A=h^{-1}\pa h$.

Integration over $\tilde{h}$ in (\ref{Gresult}) or $h$ in
(\ref{Gtilderesult}) produces $\s$-models with backgrounds given by
(\ref{OurSol}) and (\ref{DualBG}) respectively. Integration also
produces non-trivial Jacobians from changing variables, which will be
the topic of the next section.

It is interesting to investigate what happens to (\ref{WZWonD}) in the
case when $\tilde{G}$ is abelian.
If we decompose $l$ as $l=h\tilde{g}$ and introduce a lagrange
multiplier $\bar{A}$ to put the delta function in the action we get
\be{TradFterm}
 S=\int d^{2}z \left( A\tilde{\bar{J}} + \left(A\a^{-1}E^{0}
    -\tilde{J} - A\b^{t}\right)\bar{A} \right).
\eq
Now, using that in the abelian case the formulas (\ref{notation}) simplify
to ($\c$ are the coordinates on the abelian group $\tilde{G}$)
\be{AbelOmegas}
 \a=\mathbf{1} \quad \b = \c_{c}f^{c}_{ab} \equiv \c_c f^c,
\eq
and that the (abelian) currents can be written as $\tilde{J}=\pa\c$,
we can write the action as
\be{SecondTrad}
 S=\int d^{2}z \left( A\bar{\pa}\c - \pa\c\bar{A} +
   A\left(E^{0}+\c_{c}f^{c}\right)\bar{A}\right).
\eq
By partial integration we can transform this into two terms; a
constraint saying that the field strength of $A,\bar{A}$ should vanish
($\c$ becomes the lagrange multiplier) and a term
$AE_{0}\bar{A}$. Thus we recover the formulation of traditional
non-abelian duality in the form it has after gauge fixing the
coordinates on $G$ to zero. In our formalism, the ``gauge field''
$A,\bar{A}$, which is not a gauge field since we recover the
traditional formalism {\em after} gauge fixing, comes from combining a
lagrange multiplier and a field coming from the decomposition of $l$
whereas in the traditional case the $A,\bar{A}$ field comes from
gauging the isometries of the background $E_{0}$. The fact that $\b$
and $J$ were related to $\c$ in a simple manner was also important for
this equivalence to work.  This discussion might give some hints as to
why (usually) the dual background have no isometries. Only in the very
special case described above do different components combine to give
us the usual ``gauging of isometries'' description.

\section{Quantum Analysis}\label{quantum}

The determinants in (\ref{Gresult}) and (\ref{Gtilderesult}) can be
computed using standard heat kernel regularization techniques
\cite{bush,shift}. The result can be absorbed in the extra shift of
the dilaton
\be{Dilaton}
 \F^{\prime} = \F^0 + \ln\left( \det\left(E(g)\right)\right)
     -\ln\left( \det\left(E^{0}\right) \right)
\eq
and for the dual model
\be{DualDilaton}
  \F^{\prime} =
\F^0 + \ln\left( \det \left(\tilde{E}(\tilde{g})\right) \right)
\eq
Observe that these factors are normalized so that when $\tilde{G}$ is
abelian, which means that $E(g)=E^{0}$, the dilaton shift
(\ref{Dilaton}) is zero and the shift (\ref{DualDilaton}) is the same
as in the traditional non-abelian duality case. Our result thus
reproduces the result of traditional non-abelian duality. However,
when both groups are non-abelian, one has to shift the dilaton in
both models to maintain conformal invariance.

To be able to integrate out $A$ (or $\tilde{A}$)
we have to change variables in the path integral as follows:
\be{changehtilde}
\cd\tilde{h} = \cd\tilde{A}\left(\det (\pa + [\tilde{A},\cdot ])
\right)^{-1}
\eq
or
\be{changeg}
  \cd h = \cd A\left(\det (\pa + [A,\cdot ])\right)^{-1}.
\eq
Let us analyze what happens to (\ref{Gtilderesult}) using
(\ref{changeg}) (the other case being totally analogous). To avoid
gravitational anomalies and not to introduce extra contributions to
the central charge, we see that we have to include an extra factor of
$\det (\pa)$ in our starting path-integral (\ref{WZWonD}).
If we do this, the factor that we need to worry
about is
\be{ExtraFactor}
\int \cd A \frac{\det(\pa)}{\det(\pa+[A,\cdot])}
  \d\left[A-\tilde{J}\tilde{E}\a\right].
\eq
To make it more symmetrical, we integrate also over $\bar{A}$ and
include an additional delta function:
\be{SymmExtra}
\int \cd A \cd \bar{A}
   \frac{\det(\pa)\det(\bar{\pa})}
   {\det(\pa+[A,\cdot])\det(\bar{\pa}+[\bar{A},\cdot])}
  \d\left[A-\tilde{J}\tilde{E}\a\right]
  \d\left[\bar{A}\right].
\eq
The determinants in the denominator can be lifted into the action using
two extra scalar fields $\f,\f^{\star}$ transforming in the adjoint and
coadjoint representations of $G$ respectively. We thus acquire a factor
depending on $A,\bar{A}$ in the path-integral
\be{BadFact}
N(A,\bar{A})=\int \cd \f^{\star} \cd \f \exp
\left(-\frac{1}{2\p}\int d^{2}z\:
 \f^{\star}\de^{\m}\de_{\m}\f\right).
\eq
To investigate the conformal properties of $N$ we couple it to the
world sheet gravity and calculate the one-loop effective action.  The
purely gravitational parts gives a contribution to the central charge
which is cancelled by the compensating factors $\det\left(
\pa\bpa\right)$ introduced above. There is also a mixed anomaly, with
the fluctuation of the metric and a ``gauge field'' as external
fields. It gives a contribution to the trace of the energy momentum
tensor proportional to the trace of the structure constants which
cannot be absorbed in the dilaton, exactly as in \cite{AGMR,AAGY}.
This means that if $f^{a}_{ab}\neq 0$, the $\tilde{G}$ model is not
conformal, and conversely if $\tilde{f}^{ab}_{a}\neq 0$, the $G$ model
is not conformal.  One may also check, using heat-kernel methods, that
there are no additional logarithmic divergences in $N$.

Our result, depending on an ``extra'' finite, conformally invariant
factor, is the generalization of what one encounters in ordinary
abelian duality \cite{bush}. In principle $N(A,\bar A)$ could
contribute to higher order in the effective action, but this issue will
not be pursued further here.

\section{Spectators}\label{spectators}

We now return to the full problem as promised. The general $\s$-model
can be written as
\be{full}
\cs = \int d^2 z \left( \pa x^{i}F_{ik} \bar{\pa}x^{k}+
    \pa x^{i}F^{R}_{i\a} \bar{\pa}x^{\a} +
    \pa x^{\a}F^{L}_{\a i} \bar{\pa}x^{i} +
    \pa x^{\a}F_{\a\b} \bar{\pa}x^{\b}\right),
\eq
where greek indices $\a,\b\ldots$ are associated with inert
coordinates. The currents generated by the left group action
(\ref{varx}) are
\be{Spectatorcurrents}
K_{a} = \left(\pa x^{i} F_{ik} + \pa x^{\a}
   F^{L}_{\a k}\right)v^{k}_{Ra}, \nonumber\\
\bar{K}_{a} = v^{k}_{Ra}\left(F_{ki}\bar{\pa}x^{i} +
   F^{R}_{k\a}\bar{\pa}x^{\a}\right).
\eq
If we want them to obey non-commutative conservation laws in the same
way the currents (\ref{currents}) do, we also have to let
$F^{L},F^{R}$ and $F_{\a\b}$ transform under the left action of the
group $G$. The following equations (similar to (\ref{lieeqn})) should
be satisfied:
\be{SpecLie}
\cl_{v_{Rc}}(F_{ik})=F_{ij}v^{j}_{Ra}
   \tilde{f}^{ab}_{c}v^{l}_{Rb}F_{lk},\qquad
\cl_{v_{Rc}}(F^{L}_{\a i})=F^{L}_{\a k}v^{k}_{Ra}
   \tilde{f}^{ab}_{c}v^{l}_{Rb}F_{li},\\
\cl_{v_{Rc}}(F^{R}_{i\a})=F_{ik}v^{k}_{Ra}\tilde{f}^{ab}_{c}
       v^{l}_{Rb}F^{R}_{l\a},\qquad
\cl_{v_{Rc}}(F_{\a\b})=F^{L}_{\a i}v^{i}_{Ra}\tilde{f}^{ab}_{c}
  v^{k}_{Rb}F^{R}_{k\b}.\nonumber
\eq

To get the full solution, we use (\ref{classequiv}) as a guide:
let $E^{0}$ depend on the spectator coordinates and include
additional matrices $F^{L}_{\a a}(x^{\a})$,
$F^{R}_{a\a}(x^{\a})$, $F_{\a\b}(x^{\a})$ that couple to the
spectator coordinates in a natural way:
\be{Daction}
\cs &=& I[ l_{1} l_{2} ] - \int d^2 z\: \left[ \:
\langle l_{1}^{-1}\pa l_{1}, \tilde{T}^{a}\rangle E^{0}_{ab}
\langle\tilde{T}^{b}, \bar{\pa}l_{2}l_{2}^{-1}\rangle \right. \\
&-& \left. \langle l_{1}^{-1}\pa l_{1},
\tilde{T}^{a}\rangle F^{R}_{a\a} \bar{\pa}x^{\a}
  + \pa x^{\a} F^{L}_{\a b}\langle\tilde{T}^{b},
                \bar{\pa}l_{2}l_{2}^{-1}\rangle -
  \pa x^{\a}F_{\a\b}\bar{\pa}x^{\b} \:\right] \nonumber
\eq
Upon integrating over $l_{2}\in G$ (ignoring Jacobians in analogy
to the procedure followed in section \ref{path}), we obtain
\be{Fullspecaction}
\cz &=& \int \cd l \det\left(\pa\right) \: \d\left[\langle l^{-1}\pa l,
\tilde{T}^{b}\rangle E^{0}_{ba}
   +\pa x^{\a}F^{L}_{\a a} - \langle l^{-1}\pa l,
      \tilde{T}_{a}\rangle\right] \times\nonumber\\
 & &\exp\left[-I[l]-\int d^2 z\:\left(\langle l^{-1}\pa l,
  \tilde{T}^{a}\rangle F^{R}_{a\a}\bar{\pa}x^{\a} + \pa x^{\a}F_{\a\b}
     \bar{\pa}x^{\b}\right)\right],
\eq
where we have included the extra $\det\left(\pa\right)$ discussed in
the previous section.

We can now take (\ref{Fullspecaction}) as our
starting point for the dualization process and, by repeating the same
arguments, we end up with an action defined on $G$:
\be{GspecAction}
\cs &=& \int d^2 z\:\left[ JE\bar{J} + \pa x F^{L}E_{0}^{-1}E\bar{J} +
  JEE_{0}^{-1}F^{R}\bar{\pa}x \nonumber \right. \\
&+& \left. \pa x ( F - F^{L}E_{0}^{-1}F^{R}
  + F^{L}E_{0}^{-1}EE_{0}^{-1}F^{R})\bar{\pa}x \right] ,
\eq
or on $\tilde{G}$:
\be{GtildeSpecAction}
\tilde{\cs} = \int d^2 z\:\left[\,
\left( \tilde{J}-\pa x F^{L}\right)\tilde{E}\left(\tilde{\bar{J}}
  + F^{R}\bar{\pa}x\right) + \pa x F \bar{\pa} x \:\right] ,
\eq
where $J=g^{-1}\pa g$ and $\tilde{J}=\tilde{g}^{-1}\pa\tilde{g}$.

Buscher-type rules that generalize the known rules for abelian
\cite{bush} and traditional non-abelian \cite{que,AGMR} dualities
can be deduced from the above equations\footnote{The formulas for
general Poisson-Lie dualizeable backgrounds (as well as the
corresponding Buscher rules) were first given in
\cite{sever}. However, by using currents that transform under the
action of the group instead of using invariant currents, these
formulas, although correct, differ from the formulas of standard
non-abelian duality \cite{que,AGMR} by a similarity transformation, as
was explained in section \ref{gensolution}.}:
\be{Brules}
& &\left(E^{-1} - \m\n\right)^{-1}
       = \tilde{E}^{-1} - \b\a = E^0(x^\a) \nonumber \\
& & E^0 E^{-1} \cf^R = \tilde{E}^{-1}\tilde{\cf}^R = F^R(x^\a)
\nonumber \\
& &\cf^L E^{-1} E^0 = -\tilde{\cf}^L \tilde{E}^{-1} = F^L(x^\a) \\
& & \cf + \cf^L E^{-1} \left( E^0 E^{-1} - \mathbf{1} \right)\cf^R =
\tilde{\cf} - \tilde{\cf}^L \tilde{E}^{-1}\tilde{\cf}^R =
F(x^\a) \nonumber\\
& & \F + \ln\det E^0 - \ln\det E  = \tilde{\F} - \ln\det\tilde{E} =
\F^0 (x^\a)\, , \nonumber
\eq
where we have rewritten (\ref{GspecAction}) as
\be{explain}
\cs = \int d^2 z\:\left[ JE\bar{J} + \pa x \cf^L\bar{J} +
J\cf^R\bpa x + \pa x \cf \bpa x \right]
\eq
and introduced $\tilde{\cf^R}$, $\tilde{\cf^L}$,
$\tilde{\cf}$ similarly for (\ref{GtildeSpecAction}).
The method for applying Poisson-Lie T-duality can now be stated:
\begin{enumerate}
\item
Look for a set of vector fields $v_{R}$ and a set of constants
$\tilde{f}^{ab}_{c}$ such that equations
(\ref{SpecLie}) are satisfied for the
$\s$-model background. Notice that this is the generalization of what
one usually does in ordinary duality where one looks for $v$'s that satisfy
the Killing equations ${\cl}_{v}F=0$.
\item
The $f$'s and the $\tilde{f}$'s specify the Drinfeld double
used to calculate the matrices $\a$, $\b$ of (\ref{notation}).
\item
Use the formulas (\ref{Brules}) to find the dual background.
\end{enumerate}

We can check the result by analyzing the case when $\tilde{G}$ is
abelian. In that case $E(g)=E_{0}$ and (\ref{GspecAction}) reduces to
a $\s$-model with non-abelian isometries and spectators. The form of
the dual action (\ref{GtildeSpecAction}) is unchanged and is the same
as the one given in \cite{que,AGMR}.
We also see that equations (\ref{SpecLie}) are satisfied for these
backgrounds which means that they indeed satisfy non-abelian
conservation laws.

\section{Conclusions}\label{concl}

We have presented the unambiguous path-integral derivation of
Poisson-Lie duality and the most general actions that can be related
by such duality (c.f. (\ref{GspecAction}), (\ref{GtildeSpecAction}))
leading to generalized Buscher rules (\ref{Brules}).  We have analyzed
quantum determinants and discovered the non-trivial extra dilaton
shifts (\ref{Dilaton}), (\ref{DualDilaton}) that are needed to ensure
quantum equivalence of the theories up to the one-loop order.  Since
difficulties in going beyond one loop exist already in the simplest
case of abelian duality, we do not expect to do better here.  The
trace anomaly of traditional non-abelian duality for non-semisimple
groups (whose structure constants are not traceless) is present in our
case as well. With this in mind, one might look for non-trivial, {\em
conformal} examples of Poisson-Lie T-duality. If we contract
(\ref{bialg}), assuming the structure constants of both groups to be
traceless, we get a relation $f_{ab}^c\tilde{f}^{ab}_{d}=0$ which
tells us that such examples of Poisson-Lie T-duality can be found only
for groups of dimension $D\geq 5$.

Our method is based on the ability to decompose any Drinfeld double
element as a product of two group elements:
$l = \tilde{h}g = h\tilde{g}$.
As we have already noted, this can be done only close to the identity.
It would be interesting to investigate if there are any non-trivial effects
associated with this fact.

It would also be interesting to perform a calculation similar to
\cite{conf} in the Poisson-Lie case. We intend to address this problem
in the future.
\bigskip

\noindent
{\bf Note added:} After this paper appeared on hep-th, a paper on
related issues was submitted to hep-th \cite{sil}. There the authors
find that our field $\tilde{A}$ in (\ref{Gresult}) (or $A$ in
(\ref{Gtilderesult})) has to obey an interesting non-local and
non-linear ``unit monodromy constraint''
\be{unitmonodromy}
 P \exp\int_{\gamma}\tilde{A}= \tilde{e}\qquad \qquad
\tilde A=\tilde h^{-1}\pa\tilde h,
\eq 
where P stands for path-ordered exponential and $\gamma$ is a closed
path around the string world-sheet. This should give rise to
non-perturbative effects in Poisson-Lie T-duality. However, in the
present article we are only interested in perturbative
results. Indeed, as stated above, our results are valid only to first
order in perturbation theory. The uneasy reader may imagine the field
$\tilde{A}$ as being expanded around a backgound which satisfies the
constraint (\ref{unitmonodromy}) leaving an unconstrained integration
over infinitesimal fluctuations around the background. Of course, the
background field drops out of the calculation.

Another way to look at this problem is to observe that
(\ref{unitmonodromy}) is just the well-known identity stating that a
pure gauge field is flat
\be{ident}
 P \exp\int_{\gamma}\tilde{h}^{-1}d\tilde{h}\equiv \tilde{e},
\eq
but written in light cone gauge. Thus the obstruction to making our
statements globally valid is the same as the obstruction to imposing
light cone gauge globally. In particular, this should not be a problem
on any contractible patch on the world-sheet.

We would like to thank the referees for prompting us to
elaborate on this issue.

\section*{Acknowledgments}

It is a great pleasure to thank Martin Ro\v{c}ek for stimulating
discussions and inspiration.  We are also grateful to Jan de Boer and
Ulf Lindstr\"{o}m for carefully reading this manuscript.

\end{document}